**Title:** Cryogen-free low-temperature photoemission electron microscopy for high-resolution nondestructive imaging of electronic phases

**Authors**
Chen Wang,[1,2]† Shaoshan Wang,[1,2]† Chuan Guo,[3,4]† Chengjian Yu,[1,2] Qi Fu,[1,2] Xiaopeng Xie,[1,2] Changxi Zheng,[1,2]*

**Affiliations**
[1]Research Center for Industries of the Future, Westlake University, Hangzhou 310024, China.
[2]Department of Physics, School of Science, Westlake University, Hangzhou 310024, China.
[3]College of Advanced Interdisciplinary Studies, National University of Defense Technology, Changsha, Hunan 410073, China.
[4]Nanhu Laser Laboratory, National University of Defense Technology, Changsha, Hunan 410073, China.

†: These authors contributed equally to this work.
*: Corresponding author. Email: zhengchangxi@westlake.edu.cn

**Abstract**
Quantum materials exhibit phases such as superconductivity at low temperatures, yet imaging their phase transition dynamics with high spatial resolution remains challenging due to conventional tools' limitations—scanning tunneling microscopy offers static snapshots, while transmission electron microscopy lacks band sensitivity. Photoemission electron microscopy (PEEM) can resolve band structures in real/reciprocal spaces rapidly, but suffering from insufficient resolution for (near)atomic-scale quantum physics due to the unstable cooling designs. Here, we developed cryogen-free low-temperature PEEM (CFLT-PEEM) achieving 21.1 K stably. CFLT-PEEM attains a record-breaking resolution of 4.48 nm without aberration correction, enabling direct visualization of surface-state distribution characteristics along individual atomic steps. The advancement lies in narrowing the segment of band structures for imaging down to 160 meV, which minimizes the chromatic aberration of PEEM. CFLT-PEEM enables rapid, nondestructive high-resolution imaging of cryogenic electronic structures, positioning it as a powerful tool for physics and beyond.

## 1. Introduction

Different from transmission electron microscopy (TEM) and scanning electron microscopy (SEM), photoemission electron microscopy (PEEM) utilizes photoexcited electrons from occupied states for imaging [1]. This unique capability makes it a powerful tool for directly visualizing electronic band structures and distributions in both reciprocal and real spaces. Beyond static imaging, PEEM is capable of imaging dynamic processes with exceptional time resolutions ranging from attoseconds to milliseconds, including femtosecond and picosecond timescales [2,3]. Driven by several significant technological advancements, PEEM is now applicable to a wide range of studies, encompassing electronic phase transition, magnetism, ferroelectricity, semiconductor devices, plasmonics, chemical reactions, elemental composition analysis, and biological systems [4-11].

Despite its diverse applications, PEEM has primarily been used to investigate phenomena at or above room temperature [12]. Low-temperature phenomena, such as electronic phase transitions in quantum materials at cryogenic temperatures—particularly within the liquid helium range—remain underexplored using PEEM. This is primarily due to the challenges associated with maintaining stable and low temperatures for extended periods using liquid helium flow or bath, which can negatively affect spatial resolution and the study of phase transition dynamics [13-16]. Consequently, the spatial resolution of aberration-corrected PEEM equipped with an immersion lens is limited to 30 nm or worse at cryogenic temperatures, while a similar system can achieve a resolution of ~2 nm at room temperature with an aberration corrector [15,17-19].

Unlocking PEEM's high-resolution imaging capabilities at cryogenic temperatures could revolutionize the study of low-temperature quantum phenomena, especially given its rapid imaging rates and dual functionality with angle resolved photoemission spectroscopy (ARPES) for band structure characterization [20]. While low-temperature scanning tunneling microscopy (LTSTM) can image near-atomic-scale electronic structure fluctuations via quasiparticle interference, its band structure like patterns arise from quasiparticle scattering rather than direct imaging, which often leads to discrepancies with ARPES results [21]. These inconsistencies highlight the need for low-temperature PEEM with spatial resolution comparable to LTSTM, which can directly image band structures and resolve discrepancies between complementary techniques like LTSTM and ARPES.

Here, we introduce the development of a cryogen-free low-temperature PEEM (CFLT-PEEM) system, which is based on a self-developed closed-cycle cooling system integrated with multiple vibration isolation mechanisms. The implementation of a cryogen-free cooling system aims to achieve ultra-long-term temperature stability, facilitating the study of dynamic processes at low temperatures. Coupled with a self-developed cryogenic sample holder which not only has excellent thermal conductivity but also exhibits excellent electrical insulation to bias the sample at -15 kV without arcing, the system achieves a stable sample temperature of 21.1 K with only ± 3 mK fluctuation. This temperature stability is comparable to that of advanced cryogenic TEM sample holders which use continuous liquid helium flow to achieve atomic resolution [22]. Our CFLT-PEEM achieves a lateral spatial resolution between 4.48 nm and 5.78 nm, reaching the estimated instrument limit and enabling the observation of a single atomic step on the Cu(111) surface [23,24]. This step contrast may reflect the real space distribution of Bloch electrons along the atomic steps. Using a self-developed 7 eV laser, we also demonstrate high quality $\mu$-ARPES and coherent photoemission electron imaging. Notably, CFLT-PEEM enables continuous high-resolution imaging for over 9 hours without inducing irradiation damage to samples, a critical advantage over conventional electron microscopy techniques that often suffer from beam-induced degradation. The successful development of CFLT-PEEM opens up an unprecedented opportunity for studying cryogenic electronic phase transition dynamics with high spatial resolution.

## 2. Method

### 2.1 Sample preparations

The graphene samples used in this work were grown on Cu(110) and Cu(111) substrates, each measuring 5×5 mm² (HF-Kejing, 1 side polished, $\pm 2°$ crystal orientation), using the chemical vapor deposition (CVD) method. For PEEM measurements, the sample was mounted on a molybdenum sample cap and then loaded into the ultrahigh vacuum (UHV) preparation chamber, which had a background pressure of $5 \times 10^{-9}$ mbar. Before imaging, the samples were annealed at 400 ℃ for 10 hours to clean their surfaces.

### 2.2 CFLT-PEEM measurements

During imaging, the samples were loaded into the main chamber of CFLT-PEEM, which maintains a pressure of $4 \times 10^{-10}$ mbar, and thereafter biased at -15 kV. The lasers were directed onto the sample surface at an incidence angle of approximately 70° relative to the surface normal.

### 2.3 Calibrating the work function

The work function of the graphene samples was accurately calibrated by measuring the secondary electron cut-off (SECO) using a helium lamp ARPES system equipped with a hemispherical analyser [25]. Prior to measurements, the samples underwent a degassing process identical to that used for PEEM experiments (400 ℃ for 10 hours). The spectra recorded by the CFLT-PEEM system, as shown in Fig. 4 and Fig. 5, were energy-scaled based on the difference between the incident photon energy and the calibrated work function.

## 3. Results

### 3.1 Design and instrument of CFLT-PEEM

Fig. 1A indicates the overall schematic of our CFLT-PEEM system. The system mainly consists of six parts: (1) a commercialized electron microscope (FE-LEEM/PEEM P90, SPECS), (2) cryogen-free low temperature system, (3) vibration isolation frames, (4) sample mounting chamber, (5) sample preparation chamber and (6) laser units providing 7 eV and 4.66 eV photons. The LEEM/PEEM P90 is a non-aberration corrected system that provides 7.2 nm lateral resolution or worse in PEEM mode at room temperature or above [26]. To the best of our knowledge, the spatial resolution of this system at cryogenic temperatures has not yet been reported, while there has been a successful demonstration of low energy electron microscopy (LEEM) imaging on it at the temperature region of liquid helium recently [14]. In other similar PEEM systems, the low-temperature spatial resolution reported is around 30 nm or worse [15,17]. Thus, it is a significant challenge to achieve the spatial resolution limit of PEEM instrument at low temperature which has not been reported yet.

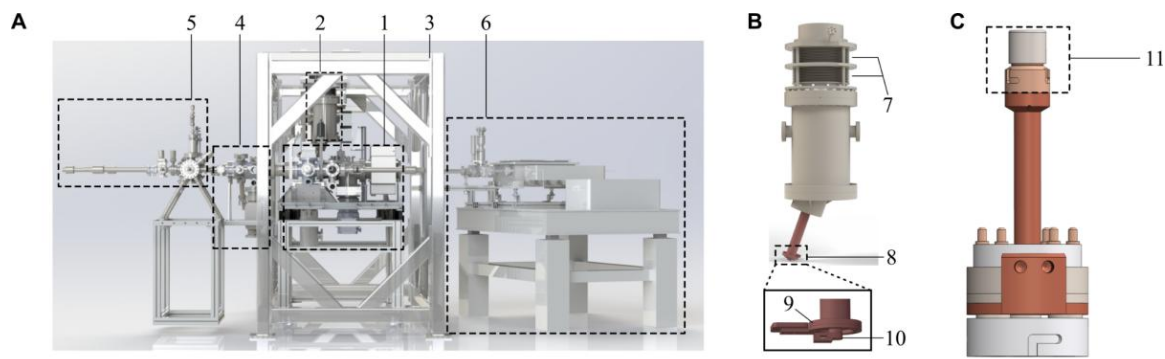

**Fig. 1. Overview of the cryogen-free low-temperature photoemission electron microscopy (CFLT-PEEM) system.** (**A**) Schematic of the CFLT-PEEM system. It mainly includes six main components: (1) commercialized LEEM/PEEM system, (2) cryogen-free cooling system, (3) vibration isolation frames, (4) sample mounting chamber, (5) sample preparation chamber and (6) vacuum ultraviolet (VUV) laser units installed on an optical table.(**B**) Cryostat and (7) pulse tube refrigerator used in CFLT-PEEM. Pressurized helium gas is liquefied in the cryostat, while (8) bellows between the pulse tube refrigerator and cryostat effectively mitigate the mechanical vibrations. The two-stage cold head comprises (9) a first-stage (90 K) and (10) a second-stage (5.3 K) for efficient sample cooling. (**C**) Layout of the self-developed low temperature sample holder for mounting (11) the sample cap.

Over the years, the development of low-temperature functionality in PEEM as well as the accompanied LEEM systems has primarily relied on wet cooling strategies, such as liquid helium flows or baths [13-17,27]. However, PEEM's relatively bulky sample and stage [15,28,29], compared to those in TEM, make it difficult to maintain low-temperature conditions stably for extended periods. This challenge limits the spatial resolution and complicates the imaging of dynamic processes, such as electronic phase transitions.

To overcome these challenges, we here successfully developed a closed-cycle cooling system tailored for PEEM imaging at low temperatures. The advantages of this system include reduced operating costs and the ability to provide a stable low-temperature environment for prolonged periods, with maintenance interruptions occurring only every few months. In our system, a pulse tube refrigerator is employed to liquefy the pressurized helium gas within a cryostat, see (7) in Fig. 1B. To isolate the mechanical vibration, the pulse tube refrigerator first mounts on the supporting frame which is fixed to the ground to direct the mechanical vibrations to the floor, see (3) in Fig. 1A [30]. Secondly, two bellows indicated as item (8) are connected to the cryostat chamber and the pulse tube refrigerator. Thirdly, the electron microscopy is placed on a vibration damping table (not shown here) which is installed inside a pit. Lastly, a self-developed sample holder (Fig. 1C) is connected to the second stage of the cryostat through a copper braid, see (10) in Fig. 1C. It should be noted that the sapphire tube mounted on the commercialized sample stage is used as the cooling shield which is connected to the 90 K first stage of the cryostat [28], see (9) in Fig. 1B. To cool the sample effectively, the sample cap (11) is screwed tightly to the sample holder rod, see Fig. 1C. The sample cap mounting and the sample preparations are carried out in the sample mounting chamber (4) and the sample preparation chamber (5) as indicated in the Fig. 1C, respectively.

To capture electronic phase transitions at low temperatures, two high-power lasers producing 7 eV and 4.66 eV photons, respectively, have been developed and installed in our electron microscopy unit. The 7 eV laser is a proven powerful tool for ARPES and PEEM, covering a certain portion of the first Brillouin zone [31,32]. Meanwhile, the 4.66 eV laser is ideal for (near-)threshold photoemission, providing photoemission electrons with minimal energy spread and divergence [19].

Fig. 2 depicts the schematic of laser sources used in CFLT-PEEM. To address the significant attenuation of 7 eV photons in the air, a vacuum chamber filled with nitrogen encapsulates related optical components and subsequent optical path. A commercial pulse laser (Coherent Paladin Compact 355-4000) with a central wavelength of 355 nm produces approximately 15 ps pulses at a repetition rate of 120 MHz, which is frequency-doubled by a KBBF ($KBe_2BO_3F_2$) crystal to generate a second harmonic (SH) beam with photon energy of 7 eV [33]. Another fundamental beam generated from a continuous wave single frequency fiber laser with a wavelength of 1064 nm is frequency doubled by a periodically poled lithium niobate (PPLN) crystal to obtain a SH beam, which is frequency doubled again by a BBO crystal ($\beta$-$BaB_2O_4$) to generate a fourth harmonic (FH) beam with a photon energy of 4.66 eV. The HW 2 (half waveplate) and QW 1 (quarter waveplate) shown in Fig. 2, are used to adjust the polarization of the 7 eV laser, and the HW 3 and QW 2 are used to adjust the polarization of the 4.66 eV laser. Moreover, the mirror 12 is used to switch between the 7 eV or 4.66 eV laser irradiation to the sample surface with an angle of incidence at 70°.

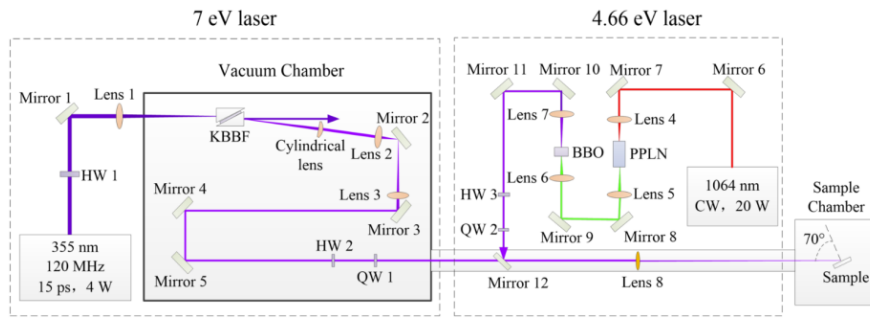

**Fig. 2. Schematic of 7 eV/4.66 eV laser source for PEEM.** The 7 eV laser is generated by frequency doubling 355 nm pulse laser via KBBF crystal. The 4.66 eV continuous-wave (CW) photons are produced via fourth-harmonic generation from a 1064 nm CW laser, employing a cascade of nonlinear optical crystals: a PPLN for second-harmonic generation (532 nm) and a BBO crystal for the subsequent fourth-harmonic process. A movable mirror (Mirror 12) switches between the two laser sources, after which the beam is focused onto the sample by a fused silica lens (Lens 8). The abbreviations of HW, QW, KBBF, BBO and PPLN stand for half waveplate, quarter waveplate, $KBe_2BO_3F_2$, β-$BaB_2O_4$ and periodically poled lithium niobate.

### 3.2 Design and performance of sample temperature control

After all the equipment developments, we first examined the temperature performance of our cryogen-free low-temperature system. During the measurement, a silicon diode temperature sensor was mounted on top of our self-developed sample holder instead of a sample cap. Since the sensor sits at the same position as the sample mounting, the temperature read by this sensor can be regarded as the sample temperature. Additionally, another temperature sensor was installed on the second stage of the cryostat. Since the sample is biased at -15 kV during PEEM imaging [34], a resistive heater was installed on the second stage of the cryostat to control its temperature and consequently control the temperature of the sample indirectly. This configuration enables precise temperature regulation via a commercial temperature controller (Lakeshore Model 335) without requiring a small heater attached directly to the sample. This setup also eliminates the need for a temperature control unit operating under high voltage bias.

Fig. 3A shows the temperature curves of the sample and the second stage of the cryostat during cooling from room temperature. As shown, the system takes around 10 hours to fully cool down. The temperatures of the second stage of cryostat and the sample can reach around 5.3 K and 21.1 K, respectively, with a maximum temperature fluctuation of ±3 mK, see Fig. 3B. By using a PID temperature controller, the temperatures of the sample and the cryostat second stage can be controlled precisely, see Fig. 3C. As shown on the right side, the temperature fluctuation amplitude of sample is less than ±7 mK in 1 min when it is stabilized at an elevated temperature. Interestingly, there is a roughly linear temperature relationship between the sample and the cryostat second stage (Fig. 3D), which means we can use this curve to deduce the sample temperature according to the temperature reading from the cryostat second stage.

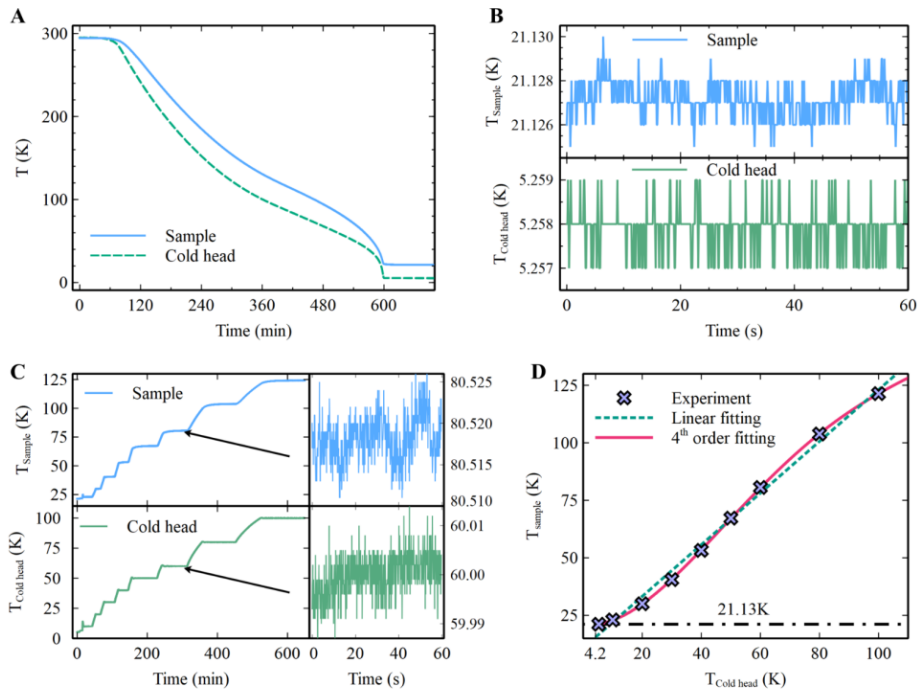

**Fig. 3. Cooling performance and temperature control of CFLT-PEEM system.** (**A**) The cooling curves of the sample and second stage starting from room temperature. The second stage of the cryostat and the sample can finally reach 5.3 K and 21.1 K within 10 hours, respectively. (**B**) The temperature fluctuations of the sample and second stage are less than ± 3 mK. (**C**) Temperature control from 21.1 K to 120 K using a PID temperature controller, and the temperature fluctuations of the sample and second stage are less than 15 mK in 1 minute. (**D**) Calibrated relationship between the second stage and sample temperature. A fourth-order polynomial fit (magenta solid line) and linear fit (green dashed line) enable precise sample temperature estimation during the measurements.

### 3.3 High resolution PEEM imaging at low temperature using 4.66 eV laser

The LEEM/PEEM P90 has a 90 degree magnetic prism which can serve as a simple energy filter with an energy resolution of 160 meV, see Supplementary Fig. S1A [35]. Fig. 4A shows the angular distributions of total photoemission electrons (ADTPE) with different energies. The ADTPE pattern was obtained from graphene/Cu(111) surface without using the energy slit, see Supplementary Fig. S2. Upon inserting the energy slit, we obtained a slice of the (k, E) spectrum, see top panel of Fig. 4B and Supplementary Fig. S1B. The detailed mechanism of forming these electron patterns in momentum space is given in Supplementary Fig. S2. The central bright spot shown in Fig. 4A corresponds to the surface state of Cu(111), as identified from other ARPES results [36].

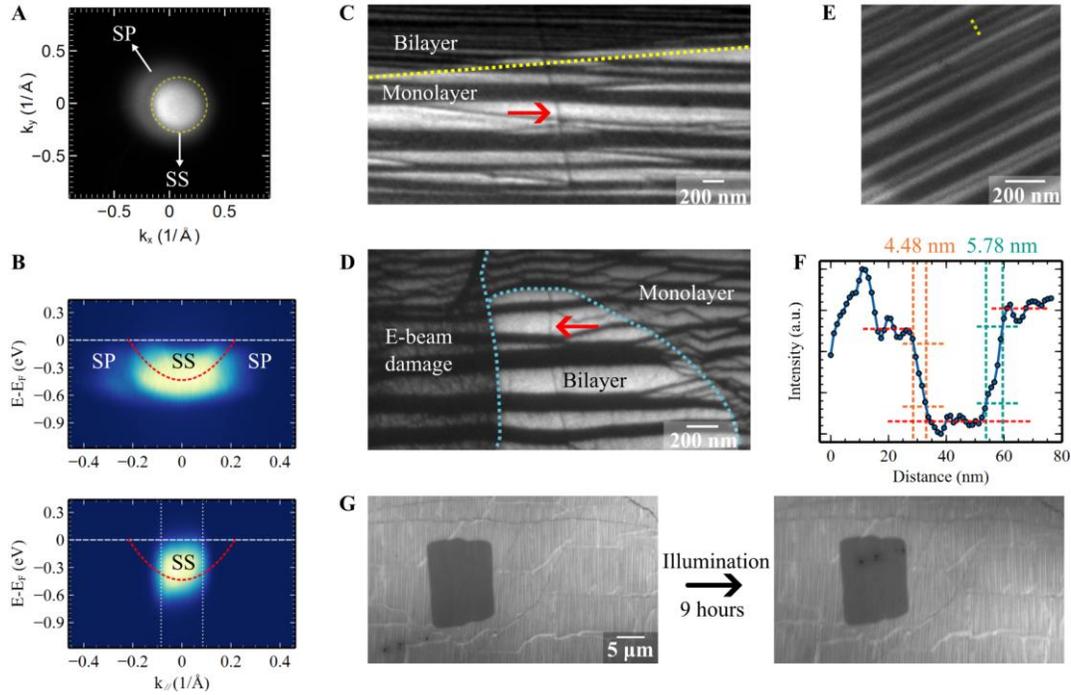

**Fig. 4. High-resolution imaging results taken with 4.66 eV laser on graphene/Cu(111) surface.** (**A**) Angular distribution of total photoemission electron (ADTPE) clearly distinguishing secondary photoemission (SP) and the surface state (SS). (**B**) Top: μ-ARPES acquired by positioning the energy slit (1 μm width) at the center of the ADTPE. The central brighter features display a nearly parabolic dispersion, consistent with the SS of Cu(111). Bottom: Energy and momentum filtering via a contrast aperture. Red dashed lines: theoretical Cu(111) surface state dispersion. Two vertical dotted lines represent the contrast aperture boundary fitted from the full width of half maximum (FWHM), while the horizontal dashed lines relate to the Fermi level. (**C**) and (**D**) Energy-filtered PEEM and LEEM images reveal consistent structural features. Atomic steps (indicated by red arrows) are clearly resolved in both PEEM and LEEM images; (**E**) High-resolution PEEM image by zooming into the bilayer region as shown in panel **C**. (**F**) Resolution analysis along the yellow dashed line indicated in e, yields spatial resolution of 4.48 nm, obtained using 16%-84% criterion, and 5.78 nm in adjacent regions. (**G**) The sample showed no illumination damage after 9 hours of continuous experiments, demonstrating the nondestructive nature of the imaging technique. Minor image shifts on the screen are attributed to electron optics adjustments during the experiments.

By sliding both the contrast aperture and energy slit inward, we selectively utilize Bloch electrons from the Cu(111) surface state for PEEM imaging, achieving an electron beam divergence of 1 mrad and a reduced energy spread. Owing to the resolution limit of the magnetic-prism-based energy analyzer in our PEEM system, we estimate the ultimate energy spread of photoemission electrons confined by the contrast aperture using the calculated band dispersion of the Cu(111) surface state (see dashed red curves in Fig. 4B). The band minimum of the Cu(111) surface state is located at -435 meV relative to the Fermi level, consistent with literature report [37]. In the bottom panel of Fig. 4B, the selected surface state segment confined by the contrast aperture is marked by two vertical dotted white lines, defined by the half-maximum slopes of the spectral peak to delineate the aperture boundary. Notably, precise alignment of the contrast aperture center with the target signal region is challenging due to operational limitations, resulting in a sampled signal area smaller than the nominal aperture diameter. Consequently, based on the calculated Cu(111) band dispersion and the aperture-defined selection range, the ultimate energy spread of electrons in Fig. 4B (bottom panel) reaches 67 meV. Meanwhile, the maximum spread accounting for the entire surface state is 435 meV, excluding weaker secondary photoemission contributions from graphene. Therefore, the energy spread utilized in our imaging system can be estimated at 160 meV, constrained by the 1 $\mu$m energy slit's resolution limit.

Fig. 4C shows the PEEM image of graphene/Cu(111) by selecting its surface state for imaging. A distinct dark line, indicated by an arrow, is observed on the surface, which is interpreted as a single atomic step of Cu. The room temperature LEEM image taken from another similar region on the same sample is shown in Fig. 4B and a similar atomic step is observed. To our knowledge, this is the first time atomic step contrast has been observed in PEEM. We can completely eliminate the possibility of step adsorption causing contrast, as the Cu (111) surface is covered by a monolayer graphene film. It is known that step contrast in LEEM is phase contrast caused by the electron interference from both sides of a step [38]. However, we cannot exclusively determine whether the step contrast in PEEM is phase contrast or not at this stage owing to the lack of interference fringes in our images. This is likely due to the contrast aperture selecting both the Cu(111) surface state and the overlapped secondary photoemission electrons from graphene for imaging, which can degrade the coherence of the imaging electrons [39]. Another possibility for the step contrast is the lower carrier density of the Cu (111) surface state along the atomic step compared to other regions. Further investigations are required to fully address this topic.

Upon zooming into the area covered by bilayer graphene, a high resolution image was obtained, see Fig. 4E. To quantitatively determine the ultimate resolution, we employed the widely used 84%-16% criterion, which defines resolution as the lateral distance between 16% and 84% of the maximum intensity [19,40]. By analysing the line profile along the yellow line shown in Fig. 4E, we determined the ultimate spatial resolution of our CFLT-PEEM to be between 4.48 nm and 5.78 nm (see Fig. 4F). To our knowledge, this is the best spatial resolution achieved by PEEM at low temperatures, with almost an order of magnitude improvement [15,17]. The resolution reaches the estimated theoretical limit of non-aberration corrected PEEM (see Supplementary Fig. S3B) and close to the value of aberration-corrected PEEM [18,19,23,24]. Particularly, the resolution simulated with 67 meV energy spread is 4.4 nm (see Supplementary Fig. S3), well fitting the experimental value of 4.48 nm. We attribute the exceptional performance of CFLT-PEEM in terms of spatial resolution to the improved photoemission electrons at low temperature and the high stability of our system. Our results bolster the confidence in the development of aberration-corrected PEEM, with the potential for spatial resolution reaching 1 nm.

In addition to improved spatial resolution, another important characteristic of CFLT-PEEM is its non-destructive imaging capability, which allows for long-term monitoring of electronic phase transition dynamics. Electron beam damage to samples is a well-known issue in other electron microscopy techniques, such as TEM, SEM (Supplementary Fig. S4), and even LEEM (Fig. 4E) [14,41,42]. However, as shown in Fig. 4G, no damage is observed in the region under a 4.66 eV focused laser beam irradiation with adequate intensity for high-resolution imaging, even after 9 hours of exposure. This is a critical factor that ensures the successful application of CFLT-PEEM to the study of electronic phase dynamics.

### 3.4 μ-ARPES excited by a 7 eV laser
Furthermore, a 7 eV laser excitation is employed for ARPES measurements, as it can cover a broader momentum space compared to the 4.66 eV laser. A series of ($E_{kin}$, k) spectra can be acquired by scanning the energy slit located at the back side of the transfer lens, as shown in Supplementary Fig. S1 [35]. From these spectra, the three-dimensional (3D) electronic structure dataset can be reconstructed, see Fig. 5A. Fig. 5B shows (E, k) dispersion extracted from the reconstructed dataset along the $\Gamma - Y$ directions of Cu(110). Fig. 5C indicates the energy resolved ($k_x$, $k_y$) maps in energy series. The surface state of Cu(110) is observed at $Y$ point. In addition to the central ring, there are two extra ring-like features around $Y$ point, which is consistent with the ARPES results taken from bared Cu(110) surface [43]. It should be noted that the pattern cutting at -2.583 eV (Fig. 5C) is the secondary photoemission electrons with dispersion [44,45]. This is because the same pattern is observed when using a 4.66 eV laser for excitation, as shown in Supplementary Fig. S5. This means that the same photoemission electron patterns consistently reside at the bottom of the free electron parabola, which is the characteristic of secondary electrons [44-46].

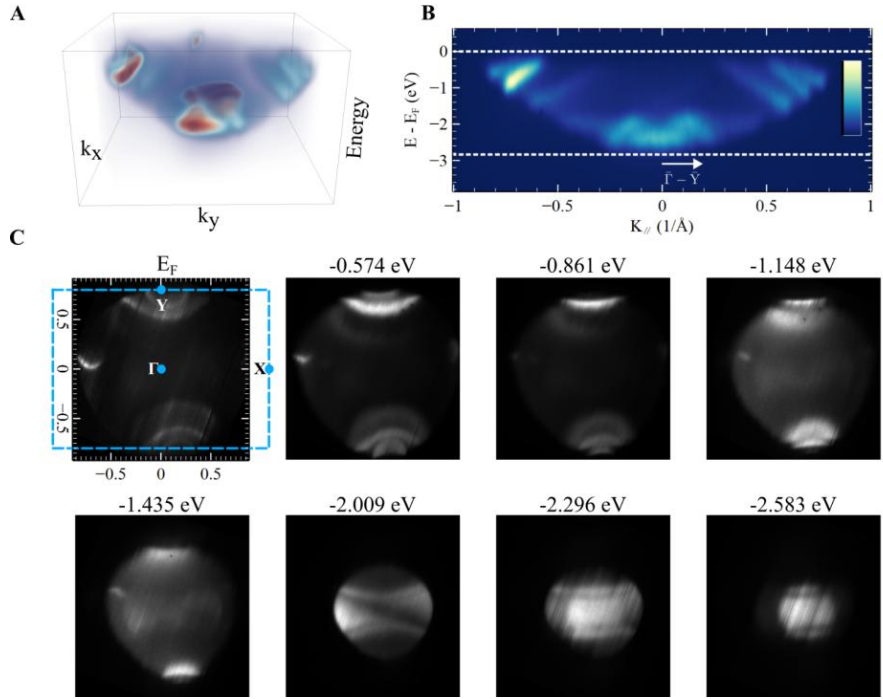

**Fig. 5. Investigation of electronic structure of graphene/Cu(110) using 7 eV photons.** (**A**) Reconstructed 3D electronic structure dataset, generated by aligning the 2D slices. (**B**) Momentum-energy (k-E) dispersion along the Γ-Y symmetry direction of Cu(110). (**C**) Energy resolved ($k_x$, $k_y$) maps at selected binding energies, extracted from the 3D dataset. Surface states (SS) of Cu(110) located Y points dominate at higher energy, while a disk with two curly stripes feature at -2.583 eV (bottom right panel) corresponds to SP signal.

### 3.5  Signature of coherent photoemission electrons

By using a contrast aperture, we can selectively isolate Bloch electrons of interest from the band structure for real-space imaging to observe their distributions at the nanoscale. Fig. 6A indicates the ADTPE pattern of Cu(110) with different energies excited by a 7 eV laser. The PEEM image formed by the total photoemission electrons is shown in Fig. 6B. As shown, the image contrast is weak and blurred, owing to the large energy spread ($\approx$ 2.8 eV) and large emission angle distribution ($\approx$ 13.51 mrad) of the total photoemission electrons. Positioning the contrast aperture at the $\Gamma$ point (the blue circle shown in Fig. 6A) for bright-field PEEM imaging can confine the emission angle as well as the energy spread of the photoemission electrons, thus yielding an image with enhanced contrast and spatial resolution (see Fig. 6C). The selected photoemission electrons are primarily secondary electrons, as mentioned above. As shown in Fig. 6C, the white straight lines and the bold white curves are atomic steps of copper and the graphene wrinkles, respectively. The bright contrast is due to the enhanced yield of secondary electrons at steps and protrusions, similar to observations in SEM [46]. Meanwhile, the dark band shown in Fig. 6C is the bilayer graphene grown on top. The contrast mechanism of bilayer graphene requires further investigations.

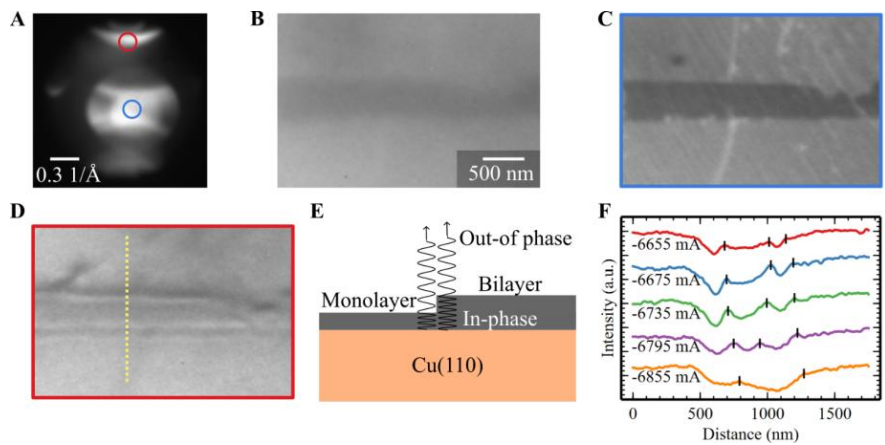

**Fig. 6. The PEEM imaging taken with 7 eV laser while selecting different angular distribution.** (**A**) Angular distribution of total photoemission electron from graphene/Cu(110) sample. (**B**) Imaging with all photoemission electron; (**C**) Bright field PEEM image with the contrast aperture positioned at Γ points, which dramatically enhances the spatial clarity compared with panel **B**. (**D**) PEEM image obtained by positioning the contrast aperture to the SS indicated by red circle in panel **A**. Two bright stripes were observed near the edge of bilayer/monolayer interface. (**E**) Proposed mechanism: Highly coherent 7 eV photons excite the Bloch electrons in Cu(110), with an additional phase difference introduced by graphene thickness variation at monolayer/bilayer edge, generating interference rings. (**F**) Evolution of line profiles (yellow dashed line in **D**) under varying objective lens focus. Stripes will move closer and eventually emerge together, confirming the coherence of interference contrast.

Furthermore, when we position the contrast aperture at the location indicated by the red circle in Fig. 6A to select the Bloch electrons of Cu(110), interference fringes are observed at the edges of the bilayer graphene ribbon, as shown in Fig. 6D. This is likely due to the photoemission electrons of the Cu(110) surface state interfering with each other after passing through the bilayer and monolayer graphene regions, as depicted in Fig. 6E. Notably, the fringes widen as the defocus increases, as seen in Fig. 6F, providing strong evidence of achieving coherent photoemission in CFLT-PEEM. It should be noted that no step contrast is observed in Fig. 6D, likely due to the larger energy spread and tilted emission angle of the surface state excited by the 7 eV laser. The quality of these photoemitted electrons is lower than that of the electrons emitted from the Cu(111) surface state through threshold photoemission.

### 3.6 Imaging at temperature regime of liquid helium

To avoid the strong surface absorption at the temperature of liquid helium region, the above PEEM measurements are carried out at the elevated temperature of 160 K. The absorption issue can be addressed soon after the installation of non-evaporable getter (NEG) pump around the sample region in the future. Thus, concise experimental results are presented here to demonstrate the capability of imaging at the temperature region of liquid helium, see Fig. 7. We fully cooled down the graphene/Cu(110) sample and acquired PEEM images using 7 eV and 4.66 eV lasers. In the low-temperature experiments, the contrast aperture and energy slit are not applied to pick up specific band structure electrons to achieve ultimate image quality. The experiment is planned to be done in the near future when better sample surface conditions can be achieved with an enhanced pumping system. Surprisingly, even without any energy filtering components, decreasing the temperature to the liquid helium regime significantly enhanced the spatial clarity compared to the 160 K condition (Fig. 6B). Furthermore, the (near-)threshold PEEM image obtained with 4.66 eV photon excitation suggests improved contrast as well, see Fig. 7B.

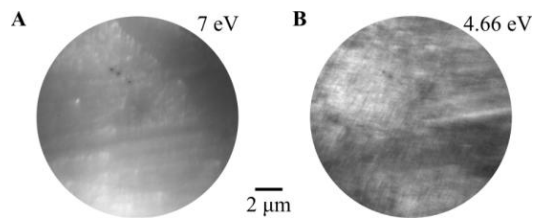

**Fig. 7. Cryogenic PEEM imaging of graphene/Cu(110) at temperature regime of liquid helium.** (**A**) PEEM image acquired with 7 eV photon excitation, revealing surface morphology contrast. (**B**) PEEM images obtained with 4.66 eV photon excitation, displaying enhanced feature resolution relative to the higher energy (7 eV) excitation. Both images demonstrate the imaging capability of CFLT-PEEM at extreme cryogenic conditions.

## 4. Discussion

In summary, we have successfully developed CFLT-PEEM equipped with a cryogen-free closed-cycle cooling system. The well-designed anti-vibration mechanisms can effectively isolate the significant mechanical vibrations generated by the pulse tube refrigerator for high-resolution imaging. The cryogen-free cooling system not only provides economic cooling costs but also offers long-term stability under low-temperature conditions. The sample temperature can reach as low as 21.1 K with a fluctuation of only ±3 mK, and it can be precisely stabilized at the desired temperature using a simple temperature controller.

Additionally, the self-developed ultraviolet lasers with photon energies of 7 eV and 4.66 eV empower the imaging capabilities, enabling detailed observations of Bloch electron distributions in both real and reciprocal spaces. Based on all these developments, our system achieves an impressive spatial resolution between 4.48 nm and 5.78 nm, reaching the theoretical resolution limit of non-aberration-corrected PEEM. More intriguingly, we observe single atomic steps and phase contrast for the first time by using Bloch electrons of surface states. Lastly, but not least, CFLT-PEEM is a new type of electron microscopy that offers nondestructive imaging capability with high spatial resolution at low temperatures, as well as a fast imaging rate. All these characteristics combined ensure high-resolution studies of electronic phase transition dynamics at low temperatures.

**Acknowledgments**
This work was supported by National Natural Science Foundation of China (Grant No. 12174319), Research Center for Industries of the Future (RCIF project No. WU2023C001) at Westlake University.

**Author contributions:** C.Z. proposed and designed the project. S.W. and C.W. developed cryogen-free low-temperature system and subsequently executed comprehensive low-temperature performance measurements. C.G. designed laser optics and optimized the spot shape. C.W. and S.W. assisted in the installation of laser units and spot tuning. C.W. and S.W. performed PEEM measurements. Q.F. conducted the SEM measurements. X.X. supplied Helium lamp ARPES measurements and precise sample work function. C.Y. conducted geometrical optics calculation. C.W. developed the code for data analysis and formatted the figures. C.Z. and C.W. analyzed the data and wrote the manuscript. All authors contributed to the discussions.

**Competing interests:** The authors declare that they have no competing interests.

**Data and materials availability:** The data that support the findings of this study are available on request from the corresponding author upon reasonable request.

# Supplementary Materials for

## Cryogen-free low-temperature photoemission electron microscopy for high-resolution nondestructive imaging of electronic phases

Chen Wang *et al.*

*Corresponding author. Email: zhengchangxi@westlake.edu.cn

**This PDF file includes:**
Supplementary Text
Figs. S1 to S5
Tables S1
References (1 to 3)

# Supplementary Text

## Ultimate resolution analysis

The theoretical resolution limit of a non-aberration-corrected (nac) PEEM system can be estimated using a geometrical optics approach [1]. By incorporating the effects of diffraction, spherical aberration (up to higher orders), and chromatic aberration, the resolution limit as a function of aperture size is given by:

$$R = [(\frac{0.61\lambda}{\alpha})^2 + (C_3\alpha^3)^2 + (C_5\alpha^5)^2 + (C_C(\frac{\Delta E}{E})\alpha)^2 + (C_{CC}(\frac{\Delta E}{E})^2\alpha)^2 + (C_{3C}(\frac{\Delta E}{E})\alpha^3)^2]^{1/2}$$

where:
- $\lambda$: wavelength of electron,
- $\alpha$: aperture angle,
- $C_3$: 3rd order spherical aberration coefficient,
- $C_5$: 5th order spherical aberration coefficient,
- $C_C$: chromatic aberration coefficient,
- $C_{CC}$: combined chromatic aberration coefficient,
- $C_{3C}$: 3rd order chromatic aberration coefficient,
- $\Delta E$: energy spread of the electron
- $E$: central kinetic energy of electron

The aberration coefficients, as detailed in Supplementary Table 1, are adopted from previous work [1]. For the high resolution image presented in Fig. 4, the energy slit and contrast aperture were optimized to exclusively select the surface state for imaging. The theoretical dispersion of Cu(111) surface state can be described by

$$\epsilon = -\epsilon_0 + \frac{\hbar^2 k^2}{2m^*}$$

which $\epsilon_0$ is set to -435 meV and $m^*$ is $0.412 m_e$ [2]. The energy spread is calculated from the aperture-defined selection range and theoretical band dispersion, yielding 67 meV. Supplementary Fig. S3 illustrates the calculated resolution as a function of acceptance angles for two energy widths $\Delta E$ (**B:** 67 meV and **C:** 200 meV) under a fixed kinetic energy of 1 eV. For the narrower energy width ($\Delta E = 67$ meV), the estimated resolution is 4.4 nm, which is consistent with the experimental results.

**Fig. S1.**

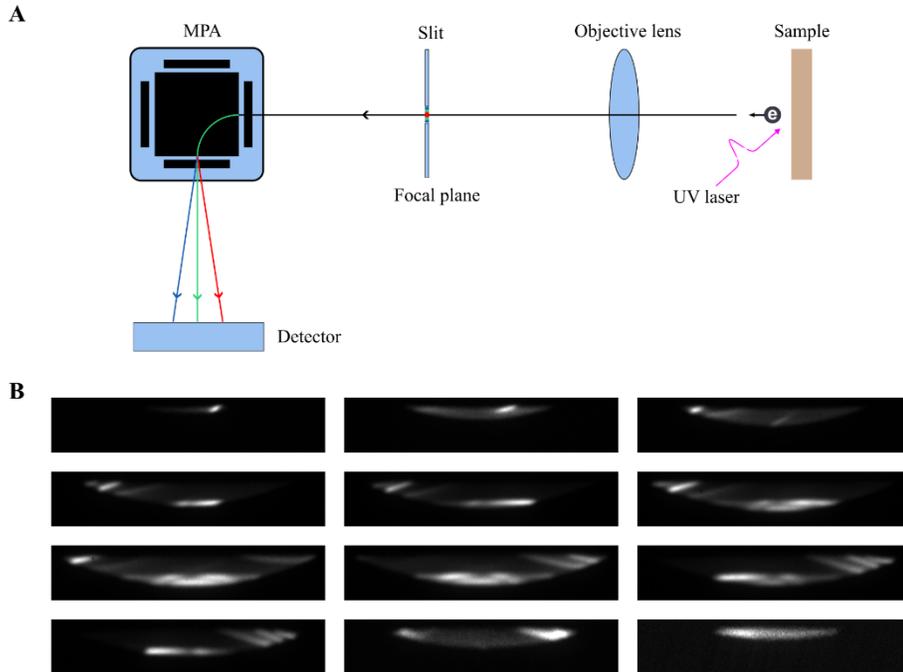

**Supplementary Fig. S1: The spectroscopic capabilities of the PEEM system. (A)** Schematic diagram of the electron optics in the PEEM system operating in the spectra mode. An external entrance slit, positioned at the back focal plane, is inserted into the electron optical path. After passing through the magnetic prism array (MPA), the electrons are dispersed based on their kinetic energy [3]. **(B)** Corresponding momentum-energy (k-E) spectra sequence (from top left to bottom right) obtained by scanning the slit through the ADTPE. Only a subset of frames is shown here, with a smaller step size used to enhance the continuity and smoothness of the 3D ARPES dataset.

**Fig. S2.**

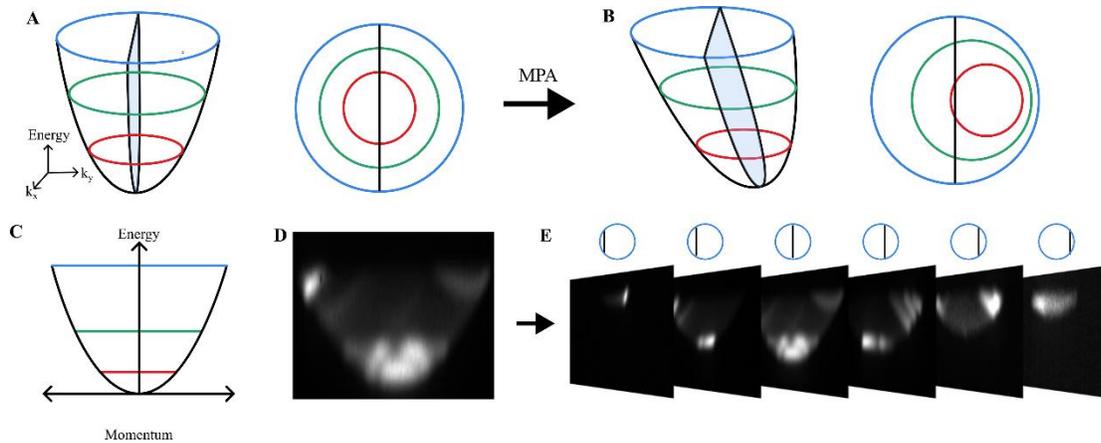

**Supplementary Fig. S2: Schematic of 3D ARPES reconstruction with in-line energy slit. (A)** Left: Full dispersion of emitted electron, where the black solid line represents the photoemission edge. Right: Top-down view of the dispersion, showing concentric circular energy contours projected onto the focal plane. **(B)** Paraboloid would be skewed after passing through the magnetic prism array, and the corresponding energy contours projected on the focal plane would also be shifted according to the energy. It induces the asymmetry in the ADTPE shown in Fig. 6A. **(C)** Momentum-energy (k-E) spectra acquired with a centered energy slit (indicated in panel B). **(D)** Experimental (k-E) spectrum of graphene/Cu(110) obtained with 7 eV photon excitation. **(E)** Sliding the energy slit across the k-space, a set of spectra can be obtained. These spectra are stacked to reconstructed the full 3D ARPES dataset.

**Fig. S3.**

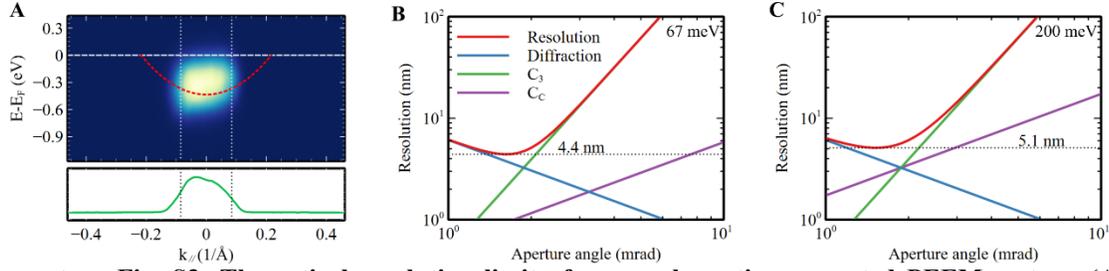

**Supplementary Fig. S3: Theoretical resolution limit of a non-aberration-corrected PEEM system. (A)** Top: Energy and momentum filtered by contrast aperture to obtain ultimate high resolution PEEM image shown in Fig. 4. Bottom: Intensity distribution obtained by integrating the data in top panel along the energy axis. Two vertical dashed lines correspond to the contrast aperture boundary. Red dotted line relates to the dispersion of Cu(111) surface state. **(B-C)** Calculated resolution as a function of acceptance angle for two different energy spreads $\Delta E$, **(B)** 67 meV and **(C)** 200 meV, at a fixed kinetic energy 1 eV. The optimal resolutions under these conditions are 4.4 nm and 5.11 nm respectively.

**Fig. S4.**

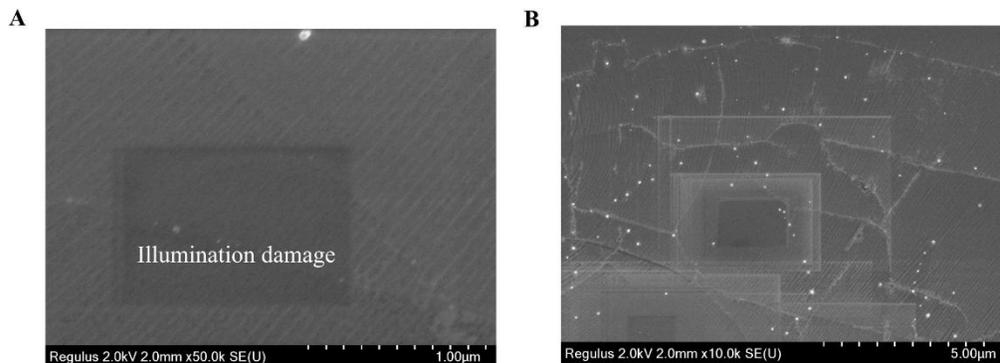

**Supplementary Fig. S4: Electron beam damage under SEM illumination.** **(A)** SEM image of graphene/Cu(111) sample at ×50 k magnification. To resolve fine features the electron beam would be focused into a small area with prolonged exposure time, resulting in significant illumination damage. The central rectangular region, illuminated for a while before, demonstrates the surface structure sensitivity to electron beam exposure. **(B)** A larger field of view (×10 k magnification) reveals that illumination damage is a common phenomenon during the SEM observation. The kinetic energy of the electron beam applied for imaging is 2 keV.

**Fig. S5.**

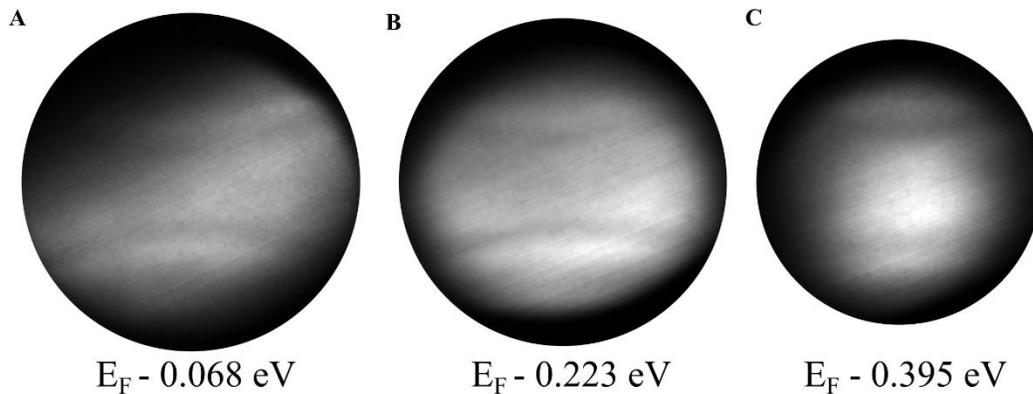

$E_F$ - 0.068 eV       $E_F$ - 0.223 eV       $E_F$ - 0.395 eV

**Supplementary Fig. S5: Energy-resolved momentum ($k_x$, $k_y$) mapping of graphene/Cu(110) using 4.66 eV laser.** Similar to the low energy side observed in Fig. 5C with 7 eV photons, the two curly stripe features also appear here. This similarity further supports the conclusion that these electrons originate from the secondary photoemission process rather than the intrinsic band structure of the material.

**Table S1.**

Aberration coefficients used in resolution calculations adopted for nac-LEEM/PEEM system with kinetic energy of 1 eV [1].

| Energy (eV) | $C_3$ | $C_5$ | $C_C$ | $C_{3C}$ | $C_{CC}$ |
|---|---|---|---|---|---|
| 1 | 0.492 | 768 | -0.13 | -1484 | 719 |

## Supplementary Reference